\documentclass[usenatbib]{mnras}
\usepackage{graphicx}
\usepackage{txfonts}

\usepackage{url}
\usepackage{graphicx}
\usepackage[hyphenbreaks]{breakurl}
\usepackage{float}
\usepackage{subfig}
\usepackage{multirow}
\usepackage{natbib}

\title[Spectral signatures of RBEs]{Searching for signatures of H $\alpha$ spicule-like features in the solar transition region}

\author[Vilangot Nhalil et al.]{Nived Vilangot Nhalil,$^{1,2}$\thanks{E-mail: Nived.Vilangot.Nhalil@armagh.ac.uk}
Juie Shetye$^{3,1}$ , J. Gerry Doyle$^{1}$ \\
$^{1}$Armagh Observatory \& Planetarium, College Hill, Armagh, BT61 9DG, N. Ireland\\
$^{2}$Astrophysics Research Centre (ARC), School of Mathematics and Physics, Queens University, Belfast, BT7 1NN, N. Ireland\\
$^3$Department of Astronomy, New Mexico State University, Las Cruces, 88001, USA }

\begin{document}
\outer\def\gtae {$\buildrel {\lower3pt\hbox{$>$}} \over 
{\lower2pt\hbox{$\sim$}} $}
\outer\def\ltae {$\buildrel {\lower3pt\hbox{$<$}} \over 
{\lower2pt\hbox{$\sim$}} $}
\newcommand{\Msun}{$M_{\odot}$}
\newcommand{\lsun}{$L_{\odot}$}
\newcommand{\Rsun}{$R_{\odot}$}
\newcommand{\solar}{${\odot}$}
\newcommand{\kep}{\sl Kepler}
\newcommand{\ktwo}{\sl K2}
\newcommand{\tess}{\sl TESS}
\newcommand{\swift}{\it Swift}
\newcommand{\Porb}{P_{\rm orb}}
\newcommand{\nuorb}{\nu_{\rm orb}}
\newcommand{\eplus}{\epsilon_+}
\newcommand{\eminus}{\epsilon_-}
\newcommand{\cd}{{\rm\ c\ d^{-1}}}
\newcommand{\MdotL}{\dot M_{\rm L1}}
\newcommand{\Mdot}{$\dot M$}
\newcommand{\Mdotsolar}{\dot{M_{\odot}} yr$^{-1}$}
\newcommand{\Ldisk}{L_{\rm disk}}
\newcommand{\src}{KIC 9202990}
\newcommand{\ergscm} {erg s$^{-1}$ cm$^{-2}$}
\newcommand{\rchi}{$\chi^{2}_{\nu}$}
\newcommand{\chisq}{$\chi^{2}$}
\newcommand{\pcmsq} {cm$^{-2}$}

\newcommand\Ion[2]{#1$\;${\scshape{\uppercase\expandafter{\romannumeral#2}}}}
\providecommand{\lum}{\ensuremath{{\cal L}}}
\providecommand{\mg}{\ensuremath{M_{\rm G}}}
\providecommand{\bcg}{\ensuremath{BC_{\rm G}}}
\providecommand{\mbolsun}{\ensuremath{M_{{\rm bol}{\odot}}}}
\providecommand{\teff}{\ensuremath{T_{\rm eff}}}

\maketitle
\begin{abstract}

New instrumental and telescopes covering the optical and ultra-violet spectral regions have revealed a range of small-scale dynamic features, many which may be related. For example, the range of spicule-like features hints towards a spectrum of features and not just two types; however, direct observational evidence in terms of tracking spicules across multiple wavelengths are needed in order to provide further insight into the dynamics of the Sun's outer atmosphere. This paper uses H $\alpha$ data obtained with the CRisp Imaging SpectroPolarimeter instrument on the Swedish 1-m Solar Telescope, and in the transition region using the Interface Region Imaging Spectrograph with the SJI 1400 \AA\ channel plus spectral data via the Si {\sc iv} 1394 \AA\ line to track spicules termed Rapid Blue-shifted Excursions (RBEs). The RBEs as seen in the H $\alpha$ blue-wing images presented here can be sub-divided into two categories; a single or multi-threaded feature. Based on the H $\alpha$ spectra, the features can be divided into events showing broadening and line core absorption, events showing broadening and line core emission, events with a pure blue shifted H $\alpha$ profile without any absorption in the red wing, broadened line profile with the absorption in the blue stronger compared to the red wing. 
From the RBE-like events which have a Si~{\sc iv} 1394~\AA\ line profile, 78\% of them show a Si~{\sc iv} line flux increase. Most of these features show a second broadened Si~{\sc iv} component which is slightly blue-shifted. 

\end{abstract}


\section{Introduction}

 Improvements to solar telescopes and their instruments has revealed a range of different small-scale dynamic features, in particular spicule-like features. Type-I spicules show a rising and declining phase, while Type-II spicules seem to lack a declining phase  as seen in Ca {\sc ii} H \citep{bart_2007}. However, they do show a declining phase in observations taken in higher temperature plasma observed with the Interface Region Imaging Spectrograph \cite[IRIS]{IRIS_paper} and Solar Dynamic Observatory Atmospheric Imaging Assembly filters \cite[AIA]{AIA_paper}, e.g. reported by \cite{Pereira_2014}, \cite{skogsrud_2015}. Sometimes they are observed as red-shifts, e.g. \cite{Bose_2021}, similar to other chromospheric down-flows \citep{Teriaca1999,Doyle2002}. Recently, \cite{Chen2022} used a 3D MHD model to show that some of these red-shifts are a result of spicules. Although other processes may be involved, e.g. pressure enhancement in the transition region, transition region brightenings unrelated to coronal emission, the boundary between cold and hot plasma, and siphon-type flows. In another study, \cite{Nived2022} showed that the number density of spicules close to the foot-points of coronal loops may be related to the amount of coronal heating.

\begin{figure*}
\centering
\includegraphics[width=0.90\textwidth]{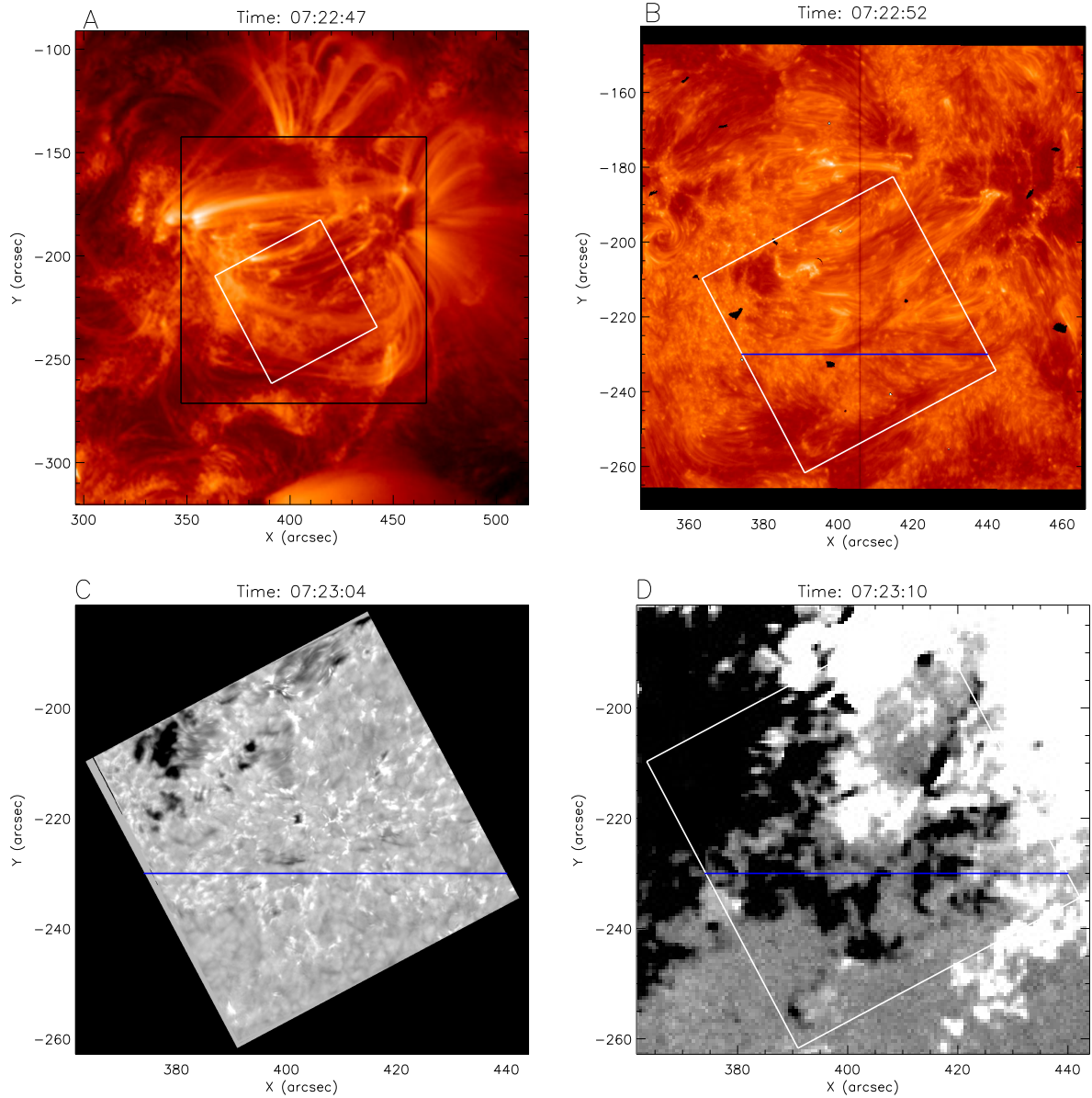}
\caption{Panel (a): AIA 171 \AA\ map of the active region. The black rectangle is the SJI FOV, while the white rhomboid is the CRISP FOV.
Panel (b): SJI 1400 \AA\ image of the active region. The white rhomboid is the CRISP FOV, while the region below the blue line is the area 
used for the detection of RBEs. Panel (c): H $\alpha$ image at --1.03 \AA. The area below the blue line marks the region of interest (the region 
used for the detection of RBEs). Panel (d): HMI LOS magnetogram with the same FOV as panel (c). The area below the blue line marks the region of interest. i.e. the region used for the detection of RBEs.}
\label{fig1} 
\end{figure*}

When observed on the disk of the Sun, spicules appear as long-elongated absorption events, often originating or present near the edge of super-granules. In the vicinity of a photospheric magnetic concentration they appear as rosettes \citep{Langangen_2008, Tsiropoula_2012}. These are usually termed as rapid-red shifted excursions \cite[RREs]{Sekse_2013b} or rapid-blue shifted excursions \cite[RBEs]{Langangen_2008,sekse_2013a}, since their appearance corresponds to red and blue shifts in the wings of the H $\alpha$ spectral line. Also, \cite{cauzzi_2008} using the Ca~{\sc ii} 8542 \AA\ line, showed that bright points, similar to those seen in Ca~{\sc ii} H{2V} and K{2V} grains, have many spicule-like structures originating from small magnetic elements, but observed in the line-cores, thus being similar to fibrils. 

\cite{Judge_2011, Judge_2012, Lipartito_2014}, report a unique observational effect related to some spicule-type events, observed in the wings of H $\alpha$ and Ca~{\sc ii} lines, that do not show any evolution, and appear suddenly in the field-of-view (FOV) of the spectrograph. If these were to be interpreted as a line-shift, we would get un-physical velocities of 1000 km s$^{-1}$. \cite{Shetye2016} and \cite{Pereira_2016} showed that these spicule-type events could be due to highly dynamic movements associated with spicules that move in and out of the narrow-band filters, used to observe different parts of the H $\alpha$ line profile moving from +35 km s$^{-1}$ to --35 km s$^{-1}$. These spicule-type events may serve as a medium for high-frequency wave transport, e.g. \cite{Shetye_2021}, thus contributing to the mass and energy transport across the solar atmosphere \citep{srivastava_2017}. \cite{Bate2022} recently showed that if even a small fraction of the wave energy 
carried by the detected transverse waves (due to spicules) was deposited as thermal energy, then it may significantly contribute to the energy requirements needed to balance the radiative losses of the chromosphere.

The range of spicule-like features hints towards a spectrum of spicule-like features and not just two types; however, direct observational evidence in terms of tracking spicules across multiple wavelengths are needed. There has been multiple efforts to track spicules across the solar atmosphere, e.g. \cite{Pereira_2014} showed that RBEs spicules undergo thermal evolution, at least to the transition region. \cite{Henriques_2016} suggested that at least 11$\%$ of RBEs could be connected to coronal counterparts. In previous work, we used imaging information to track RBEs via a tracking algorithm to identify and track them within the chromosphere using H $\alpha$ data obtained with the CRisp Imaging SpectroPolarimeter (CRISP) instrument on the Swedish 1-m Solar Telescope (SST), and in the transition region using the 
Interface Region Imaging Spectrograph (IRIS) with the SJI 1400 \AA\ channel plus spectral data via the Si {\sc iv} 1394 \AA\ line. 

This paper is organised as follows. Section~\ref{obs} provides the basic information on the observational data and in particular the co-alignment of the SST and IRIS data as this is critical to the discussion which follows. Section~\ref{methods} outlines the methodology used to study the events physically and in particular spicule detection in the different datasets plus their time tracking, plus a description of the spectral line properties using both Gaussian fitting and a moment analysis. In Section~\ref{cases}, we discuss specific examples in detail.  Finally, the discussion and conclusions are summarized in Section~\ref{dis}, where we discuss whether we have a forest of spicule-like features and not simply two types. 

\begin{figure}
\hspace*{-1cm}
\includegraphics[width=0.5\textwidth]{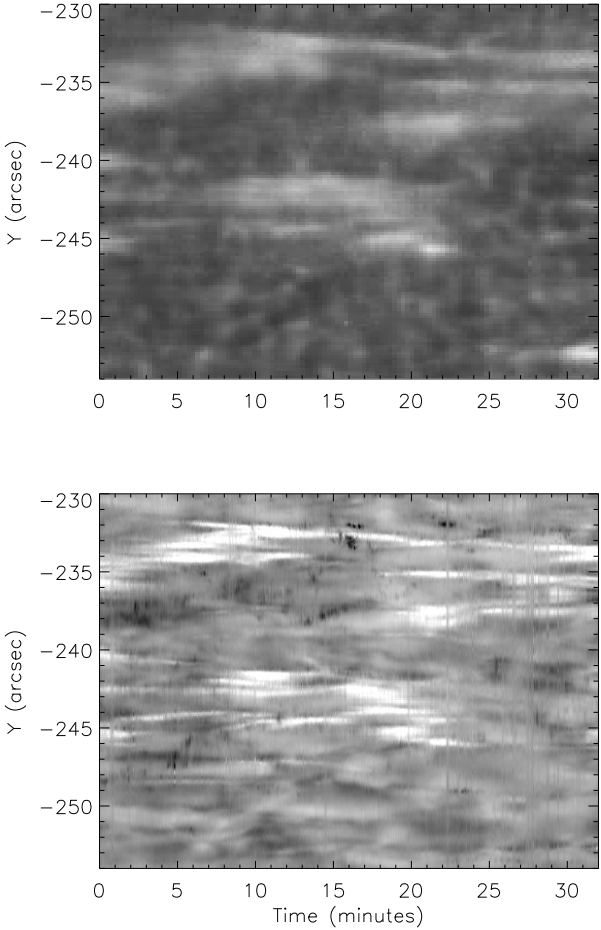}
\caption{  Panel (a): IRIS raster of the continuum at 2800 {\AA}. Panel (b): A raster taken in the H $\alpha$ wing at --1.032 {\AA}. Time 0 represents 10-Jun-2014 07:50:53.320 UT.}
\label{fig_al}
\end{figure}

\section{Observations and Data Processing} \label{obs}

\begin{figure*}
\centering
\includegraphics[width=0.96\textwidth]{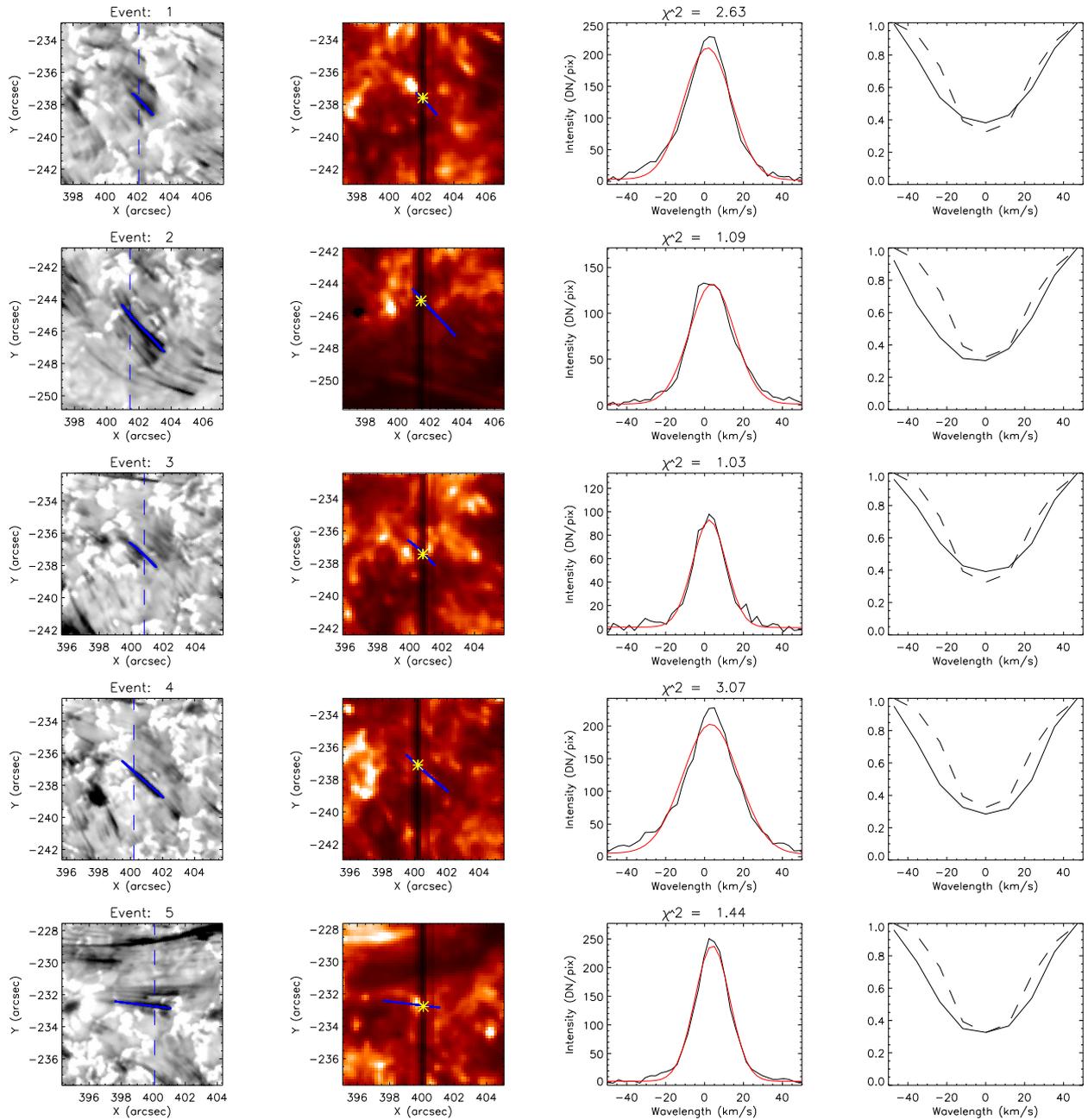}
\caption{Spectral profiles of 5 RBEs. First column: Location of the RBEs in the H $\alpha$ blue wing image, the blue dashed line shows the location of the IRIS slit. The blue curve is the axis of the RBE obtained from the OCCULT tracing code. Second column: The SJI 1400 {\AA} map corresponding to the first column. The blue line is the axis of the RBE, while the location where the RBE crosses the slit is marked in yellow. Third column: Si {\sc iv} 1394 \AA\ spectra at the location marked in yellow. Fourth column: The dashed line represents the H $\alpha$ profile of the QS region. The black thin line represents the averaged H $\alpha$ profile along the axis of the RBE. \label{fig_spectra}}
\end{figure*} 

\subsection{The data}

 \cite{Pereira2013} showed that the different properties of spicules can be reconciled given different spatial and temporal resolutions. 
To observe and trace RBEs across different atmospheric regions, we used coordinated observations from SST and IRIS, with support from SDO, obtained on 10$^{th}$ June 2014. The ground-based observations were obtained using the CRISP \citep{Scharmer_2008} instrument, with an approximate pixel scale of 0.0592$''$, that resolves around 45 km on the solar surface and has a field of view (FOV) of 60$''$ $\times$ 60$''$. After co-alignment with SDO/AIA 1600 \AA\ images  (see next section), we determined that the CRISP FOV is tilted at an angle of 62$^{\circ}$04$'$ with respect to the SDO heliocentric FOV, and centered at $xc = 40''$ and $yc = -211''$, which contained a pore and was entirely within NOAA AR 12080. Nine H $\alpha$ line-positions were sampled from the line center at H $\alpha$, with equidistant increments of $\pm$ 258 m\AA\ on both sides of the H $\alpha$ line center. The cadence of the observations is $\sim$4 s. More details about the observations can be found in \cite{Shetye2016}.  

IRIS observed active region NOAA 12080 in a sit-and-stare mode and obtained Si~{\sc iv} 1394 \AA\ spectra using the $0.33 ''$ wide slit. The cadence of IRIS observation is $\sim$16 s, which is $\sim$4 times that of SST. This difference in cadence implies that there are four SST observations corresponding to one IRIS observation. A near-perfect co-alignment between IRIS and SST is necessary to study the Si~{\sc iv} profiles associated with H $\alpha$ spicules. We performed the co-alignment using the AIA 1600 \AA\ and SJI 1400 \AA\ channels via a cross-correlation technique (see next section for details).  In Figure~\ref{fig1}, we show the context of the observations between AIA 171 \AA, the magnetogram, IRIS Si {\sc iv} 1394 \AA\ and SST H $\alpha$ images. Panel A shows the AIA 171 \AA\ FOV of the active region. The black rectangle represents the FOV of the IRIS observation. The white rhomboid is the FOV of the SST/CRISP observations. The SJI 1400 {\AA} channel map with the CRISP FOV is shown in panel B. The region below the blue line marks the region of interest in this study. This is the area used for the detection of RBEs.  

\subsection{SST and IRIS co-alignment}
 For the co-alignment of the IRIS and SST data, we first employ the cross-correlation of image SJI 1400 \AA\ and AIA 1600 {\AA} that are close in time and morphologically similar. The plate scales between image pairs are matched before starting the cross-correlation. Then we shifted the AIA image according to the offset obtained from the cross correlation algorithm. Similarly, we obtained the offset between the co-aligned AIA 1600 \AA\, and H $\alpha$ wing (at --1.032 {\AA} offset from line core) images and shifted the SST images to align with the SJI observation. We estimate that the error in the co-alignment to be better than one IRIS pixel, i.e. $\sim$0.16635$\arcsec$; we will return to see in Section 3.

To further confirm the co-alignment, we used IRIS data to compare the continuum at 2800 \AA\ (near Mg {\sc ii}) with a constructed H $\alpha$ continuum raster at H $\alpha$ --1.032 \AA. The continuum at 2800 \AA\ is formed in the upper photosphere \citep{mg_2800_carlsson}, at this wavelength magnetic features appears as bright. Similarly, at H $\alpha$ --1.032 \AA, the magnetic features also appears with an increased intensity. Therefore, we can use this wavelength to crosscheck the alignment of the IRIS slit position on the SST maps. Looking at the images, there a slight misalignment between the Mg continuum raster and the artificially rastered H $\alpha$ continuum map. We corrected this shift in both the N-S and E-W directions by maximising the cross-correlation coefficient between the images. The aligned maps are shown in Figure \ref{fig_al} which shows excellent agreement between both figures, confirming the accuracy of the co-alignment.


\section{Methodology and Analysis} \label{methods}
\subsection{Spectral properties H $\alpha$}\label{spec_avg}
The property of an event to be classed as an RBE depends on the line shift observed in the line profile of the event. The average H $\alpha$ line-profile is obtained by summing a quiet-Sun region from line-sequences as given in \cite{Shetye2016}. This quiet-Sun region is from the full FOV obtained using the CRISP instrument that is void of any major magnetic activity such as the presence of a pore. It also is a region where there are no down flows detected in the line core. The line profile defined for the event is computed across multiple locations of the RBE, and is then the average line-profile for the RBE. The pixels used for this averaging are the ones that pass through the axis of RBEs at the time frame in which the RBE appeared maximum in its length.

\begin{figure*}
\centering
\includegraphics[width=0.88\textwidth]{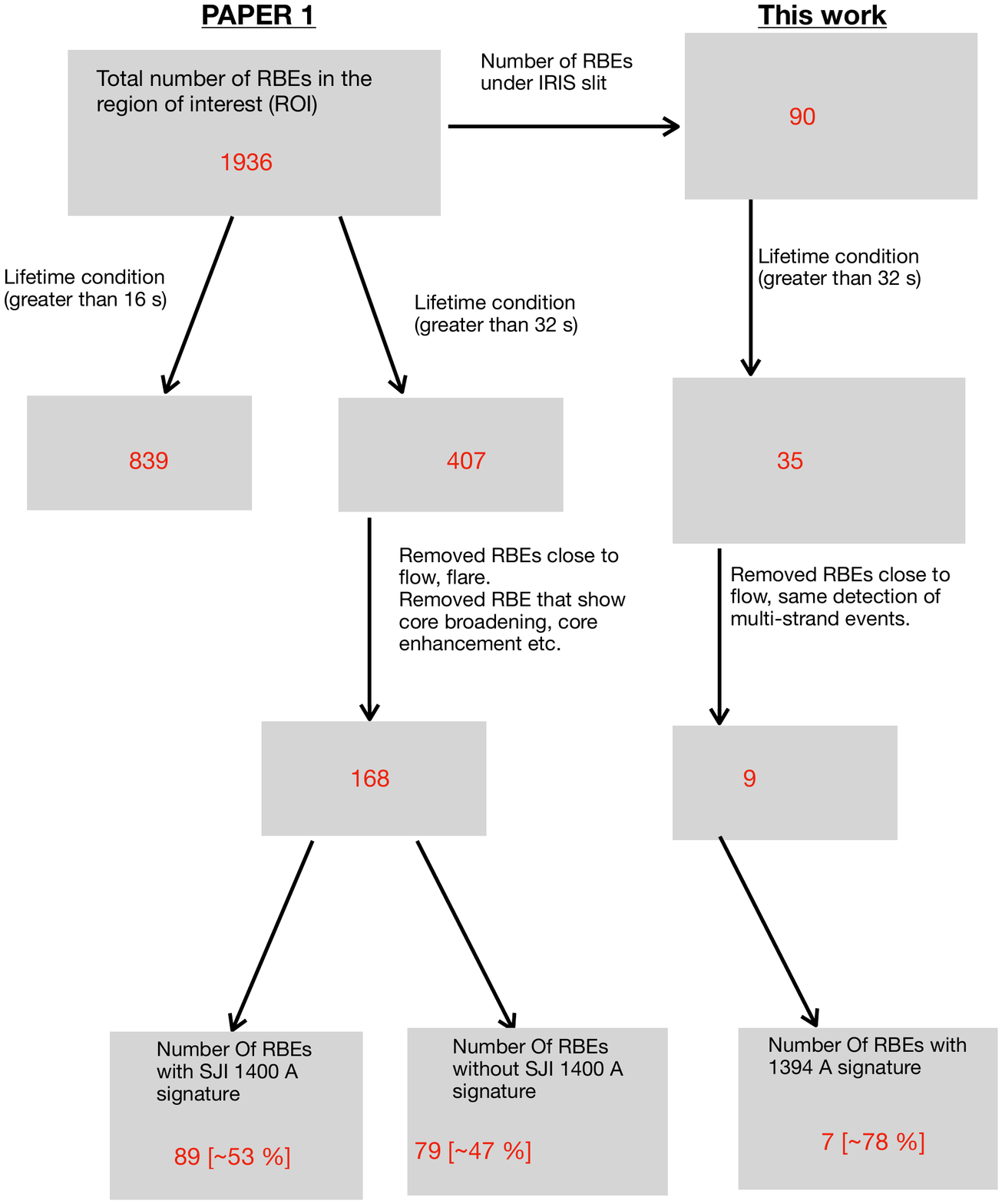}
\vspace{-3cm}
\caption{Detection summary}
\label{fig_chart}
\end{figure*}

\subsection{Detection algorithm for RBEs from H $\alpha$ images}\label{det_rbe}
We use the detection method described by \cite{Detection_paper}.  In the above paper we detected 407 RBE-like events in the H $\alpha$ data with a lifetime greater than 32 s. This number is further reduced to 168 RBE-like features since some of these features were in the proximity of a large-scale flow (i.e. we only consider the region below the blue line in Figure~\ref{fig1}. From this reduced number, only 90 of them display a clear spatio-temporal signature in the SJI 1400 \AA\ channel.

Since we are interested in whether any of these had a spectral signature in the transition region, we focus on the detection of these RBE-like features that are scanned by the IRIS slit.  All of the RBEs detected under the IRIS slit cannot be studied in Si {\sc iv} 1394 \AA\ spectra due to the limitation imposed by the cadence of the IRIS observations, i.e. 16 s. Furthermore, we add another constraint to our RBE detection. In this condition, only those RBEs are counted as detected if they occur in 2 consecutive IRIS images, i.e we have at least two IRIS spectra obtained for each detected RBE. This additional condition is satisfied by 35 RBEs. 

Lastly, we revisit the detected RBEs manually, and remove those that are under or in the vicinity of an active region flow since IRIS spectra for such events show a strong blue-shift. This procedure reduces the number of events under the slit to 18. Furthermore, we excluded those events with strong absorption in the line core of H $\alpha$. These procedures reduce the number of events under the slit to 9, which is not enough to perform a detailed statistical analysis on the connection between chromospheric and transition region emission. Therefore in this work, we will direct our interest to the different types of signatures associated with RBEs in the transition region. 

Among the 9 selected events, only 2 of them have a pure blue shifted H $\alpha$ profile without any (very weak) absorption in the red wing.  The rest of them have significant absorption in the red wing. However, the absorption in the blue is stronger compared to the red wing. Moreover, 1 event show core absorption and another 2 show core emission. The Si {\sc iv} signature associated with each type of event is studied in detail in the remaining section.

\subsection{Spectral properties in Si {\sc iv} 1394 \AA}

We derive spectral properties such as intensity, velocity and full width half maximum (FWHM) of the Si~{\sc iv} 1394 \AA\ line by fitting in the 
first instance a single Gaussian to the profile (in a later section, we also perform a moment analysis for comparison).

To calculate the Doppler shift, we first perform the wavelength calibration using the O~{\sc i} 1356 \AA\ line. Spectral lines of neutral and singly ionized species formed in the lower chromosphere are generally considered as at rest \citep{hassler_1991}. Therefore, the O~{\sc i} line can be used to obtain the reference wavelength of the Si~{\sc iv} 1394 \AA\ line by following the formula below,

\begin{equation}
    \lambda_{ref\_Si~{\sc iv}} = \lambda_{Si~{\sc iv}\_lab} - (\lambda_{O~{\sc i}\_lab}-\lambda_{O~{\sc i}\_obs}),
\end{equation}

\noindent

where $\lambda_{Si~{\sc iv}\_lab}$  is the laboratory wavelength of the Si~{\sc iv}~1394 \AA\ line (1393.755 \AA) and $\lambda_{O~{\sc i}\_lab}$ is that of the O~{\sc i} line (1355.598 \AA). The laboratory value of the wavelengths are taken from \cite{sandlin}. $\lambda_{O~{\sc i}\_obs}$ is the observed O~{\sc i} wavelength, which is obtained by fitting a single Gaussian function to the observed profile. 
Note that some of the observed Si {\sc iv} spectra cannot be fitted with a single Gaussian function as they show the presence of an additional component, which could be caused by flows and/or turbulence. For those profiles, we used a combination of two Gaussian functions (see Appendix \ref{app}).
\subsection{Detection of RBEs in Si~{\sc iv} 1394 \AA}


\begin{figure}
\centering
\includegraphics[width=0.45\textwidth]{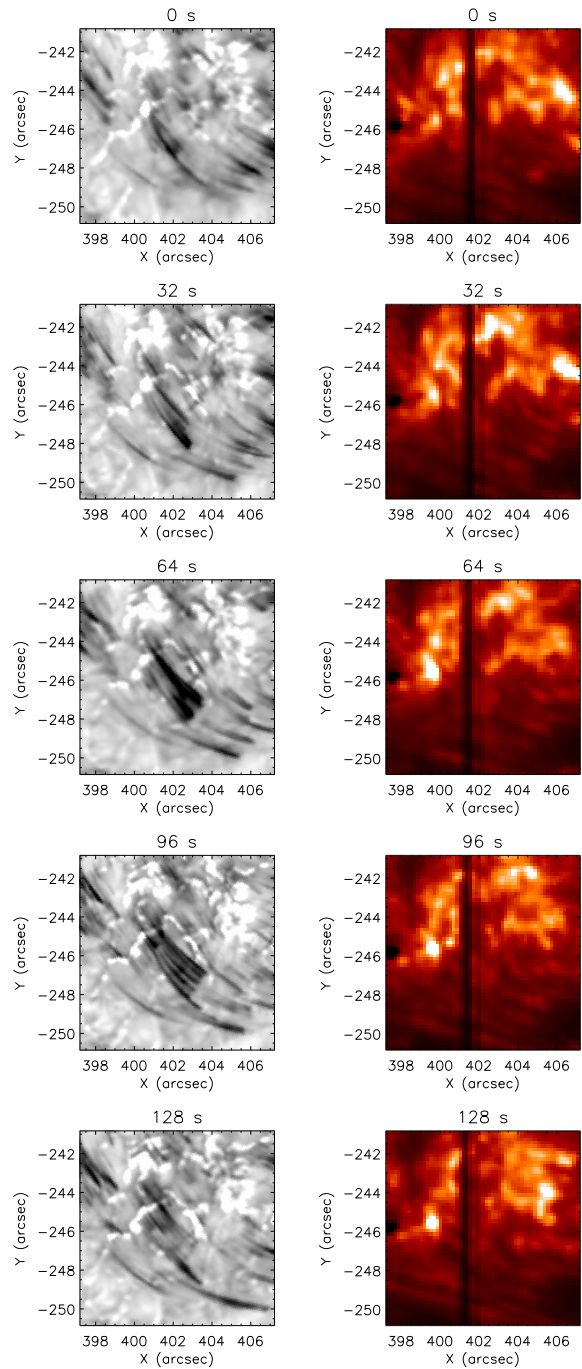}
\caption{Time evolution of event 2 in H $\alpha$ and the SJI 1400 {\AA} channel.\label{fig_t_ev}} 
\end{figure}
In the sit-and-stare observation mode, IRIS obtains the spectra at a fixed location on the Sun using a $0.33''$ wide slit. However, due to solar rotation, the location of the slit changes with time ($8.6''$ in 66 minutes). This small change in the slit location must be considered when locating the slit in the SST observations thereby identifying the RBEs that cross the slit. The location of the slit is provided in the IRIS windata structure, based on this the RBEs that cross the slit can be easily found using the coordinates of spicules obtained from the tracing algorithm (Section \ref{det_rbe}). These RBEs appear as bright jet-like features in the dataset and often appear with a brightening in the Si~{\sc iv} image. Once the crossing location is identified, we identify the RBE signature in the corresponding Si~{\sc iv} spectra. An  example of RBEs and their Si~{\sc iv} 1394 \AA\ line profiles obtained from the crossing location are shown in Figure ~\ref{fig_spectra}. Note that we cannot always expect a one-to-one match between the position of the RBEs in H $\alpha$ and Si {\sc iv} due to the large difference in the temperature of formation of these lines.
The images in the left column show the location of the RBEs in the H $\alpha$ blue wing at --35 km s$^{-1}$. This is the frame in which RBE have their maximum apparent length. The blue dashed line indicates the location of the IRIS slit, while the solid blue curve marks the axis of the RBE obtained from the OCCULT code. The second column shows the location of the RBE, represented by a blue curve in the SJI 1400 {\AA} channel. The location where the RBE crosses the IRIS slit is marked as a yellow star. The Si~{\sc iv} spectra at this location is shown in the third column.

Note that some of the Si~{\sc iv} 1394 \AA\ spectra associated with RBEs are broadened and cannot be represented by a single Gaussian function. The fourth column shows the H $\alpha$ profile of RBEs (thin black line) and the QS region (black dashed line). As mentioned in Section \ref{spec_avg}, the QS profile was computed from a section of the quiet Sun. See Appendix \ref{app} for a description of the Gaussian function. 

In Figure~\ref{fig_spectra}, Events 1 and 3 have a similar H $\alpha$ profile. They show an enhancement in intensity at the line core of H $\alpha$, with a blue-component in Si~{\sc iv} 1394 \AA. Event 2 shows a standard blue shift in the H $\alpha$ wavelength which is the spectral profile of an RBE. Event 4 shows the presence of broadening in both H $\alpha$ wings with strong absorption at the line core. Event 5 shows only a broadened line profile in the wings of H $\alpha$, however the absorption in the blue wing of H $\alpha$ is larger than in the red wing.

\section{Case Studies} \label{cases}

 To identify Si~{\sc iv} signatures associated with spicules, we construct the time-distance plot of the H $\alpha$ blue wing at --35 km s$^{-1}$ and compare it with the time-distance plot of IRIS data constructed from the Si~{\sc iv} spectra. Note that, IRIS observations are performed in a sit-and-stare mode, therefore the image constructed from the spectra can be considered as a time-distance map at the location of the slit. However, there is a shift in the IRIS slit due to solar rotation; we have taken this into account while creating the time-distance map from the SST observations.

A time-distance plot can provide information about the time evolution of the Si~{\sc iv} 1394 \AA\ intensity, velocity, and line-width at the slit location. By analysing the Si~{\sc iv} spectral properties before and after the formation of a spicule, we can detect signatures of spicules in Si~{\sc iv}, if any are present. To create a time-distance map in H $\alpha$, we first reduce the resolution of SST to that of IRIS. Then, cutting slits of width 0.33$''$ from the SST map at different time-frames. The procedure is very similar to creating an artificial raster of IRIS observation. Since the IRIS observation are in sit-and-stare mode, the rastering  procedure produces a time-distance map of the slit location.

\begin{figure*}
\centering
\includegraphics[width=0.98\textwidth]{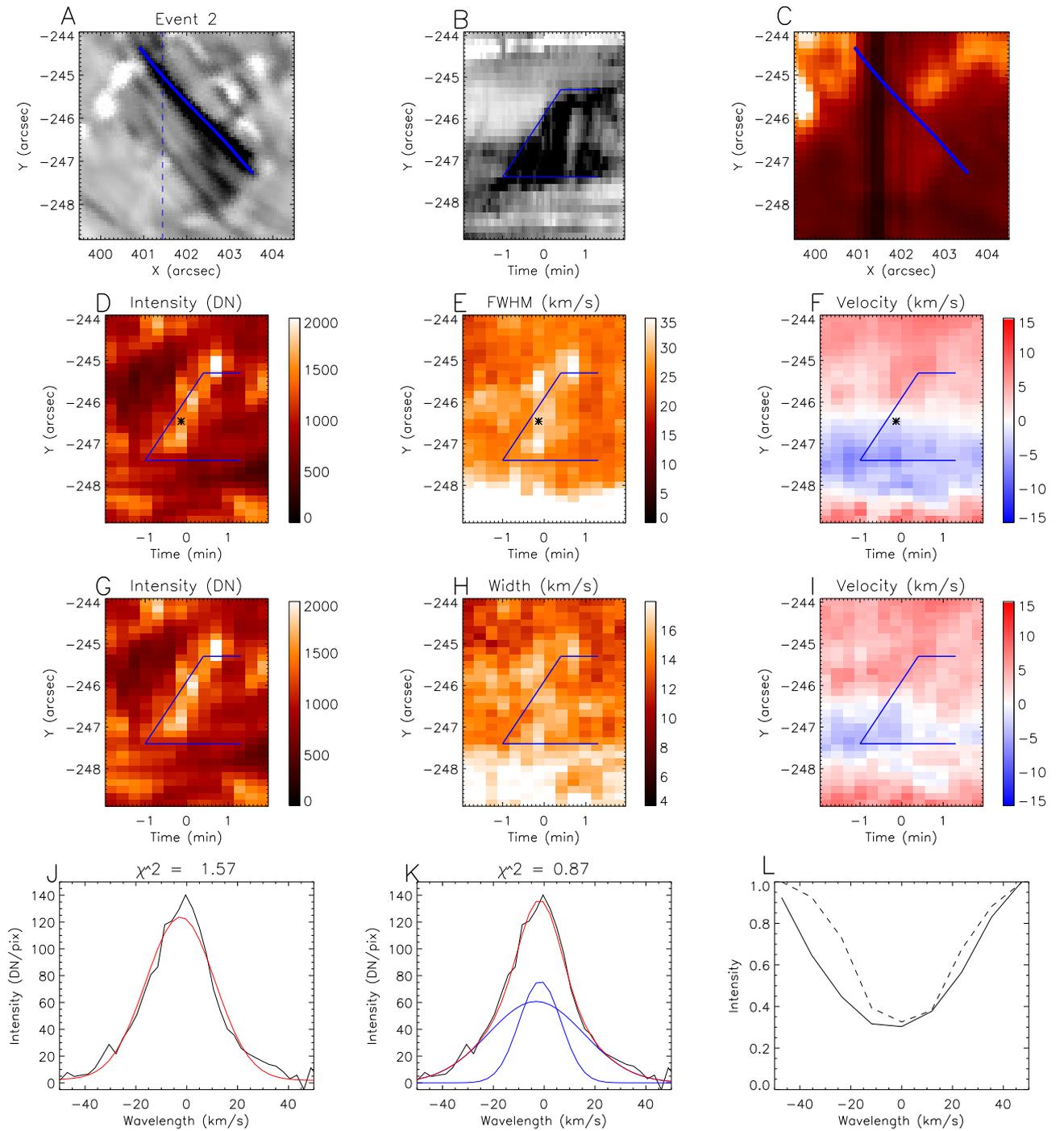}
\caption{Time evolution of event 2. Panel (A): Location of RBEs in the H $\alpha$ blue wing --35 km s$^{-1}$ image, the blue dashed line shows the location of the IRIS slit. The red curve is the axis of the RBE obtained from the OCCULT tracing code. Panel (B): The time-distance map in the H $\alpha$ blue wing. The blue curve marks the approximate location of the RBE in the time-distance plot. Panel (C): SJI 1400 \AA\  channel image of the RBE location. Panel (d): Time-distance map of Si~{\sc iv} 1394 \AA\ integrated intensity obtained by fitting a single Gaussian to the spectra. The blue curve shows the approximate location of the RBE in H $\alpha $ blue wing distance plot. Panel (E) time evolution of the FWHM width obtained from a single Guassian fit. Panel (F): Time evolution of LOS velocity. Panels (G), (H) and (I): Intensity, width, LOS velocities obtained from a moment analysis. Panel (J): An example of single Gaussian fit (red) to spectra (black) at the location marked in star in panel D: Panel K; same as panel (I), but fitted with a double Gaussian. The blue curve indicates the two Gaussian components. Panel (L): H $\alpha$ profile of the RBE (thin line) overlaid on a QS profile (dashed line). RBE profile is the averaged H $\alpha$ profile along the axis of the RBE.}
\label{fig_td_ev2}
\end{figure*}

\begin{figure*}
\centering
\includegraphics[width=0.98\textwidth]{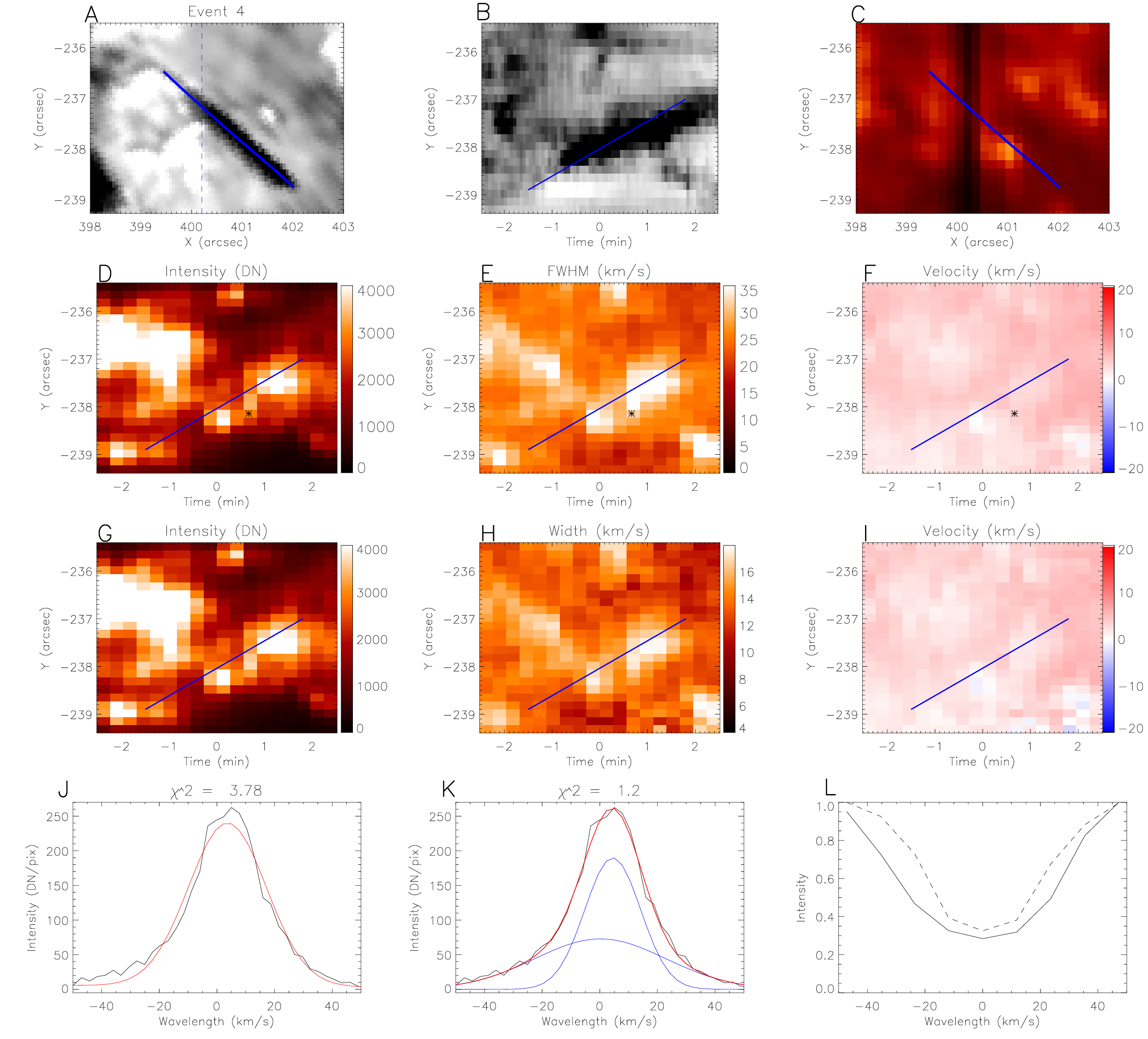}
\caption{Same as Figure \ref{fig_td_ev2} but for event 4.} \label{fig_td_ev4} 
\end{figure*}

\subsection{A multi-threaded event with RBE like profile}
Figure \ref{fig_t_ev} presents the time-evolution of an RBE (event number 2 in Figure \ref{fig_spectra}) in the H $\alpha$ blue wing at --35 km s$^{-1}$ and SJI 1400 \AA\ channel (right column). The top row shows the images taken at the starting time of the RBE. The time marked at the top left corner of the image represents the time compared to the starting time. The event shown here is an example of a group of RBEs which are close to each other, giving a multi-threaded appearance. Threads form and disappear during the lifetime of the RBE and appear to undergo transverse motion in the plane of sky. A similar observational effect was reported in \cite{Shetye_2021}. A faint thread-like structure can also be seen at the same location in the SJI 1400 \AA\ channel. However, this is not very clear due to the obstructed view caused by the presence of the IRIS slit. Furthermore, there is no evidence of bright point motion associated with this event in the SJI 1400 \AA\ channel.

The time-distance analysis of event 2 is shown in Figure \ref{fig_td_ev2}. Panel (A) shows the H $\alpha$ blue wing image of the RBE at the time frame in which the RBE appeared maximum in its length. The dashed blue line indicates the location of the IRIS slit. The coordinate of the RBEs detected (based on the OCCULT code) is shown in blue colour. Panel (B) shows the time-distance plot of the H $\alpha$ --35 km s$^{-1}$ blue wing. In the time-distance map, the RBE shows a complex time evolution due to its multi-threaded nature and the transverse motion of individual threads. It is not possible to say whether this is due to rotation or the appearance/disappearance of individual threads. 

The SJI 1400 \AA\ channel view of the region is presented in panel (C). In panels (D), (E) \& (F) we plot the time-distance plot of the Si~{\sc iv} 1394 \AA\ intensity, FWHM and LOS velocity respectively, which are derived from the single Gaussian fit to the spectra. The time-distance plots in IRIS reveal that there is a spatio-temporal enhancement in intensity and line width at the location of RBEs. Furthermore, the slope of the intensity enhancement matches very well with the slope of the RBEs in the H $\alpha$ time-distance plot. This implies that the observed intensity enhancement and the line broadening are associated with a H $\alpha$ RBE. There is a slight misalignment between the location of the spicule and its signature in Si {\sc iv}. This is expected due to the large difference in the temperature of formation of these lines. The signatures of spicules are completely absent in the Si~{\sc iv} LOS velocity measurements. However, we note that there is a large gradient in the Si~{\sc iv} velocity at the location of the RBE. The bottom panels of Figure \ref{fig_td_ev2} show the Si~{\sc iv} 1394 \AA\ and H $\alpha$ spectra of the RBE. The Si~{\sc iv} spectra is obtained from the location marked by a star on the IRIS time-distance map, while the H $\alpha$ spectrum is obtained by averaging the spectrum along the axis of the RBE.  In panels G, H and I, we show the equivalent moment plots for comparison with panels D, E and F. there is excellent agreement between the different methods.  
 
In panel J, we show the single Gaussian function (red) to the observed spectrum (black). The results have a \chisq~value of 1.57 and it fails to fit the spectrum properly. Therefore, we fit the same spectra with a double Gaussian function, with a broad component as shown in panel K. The blue curves are the two Gaussian components and the red curve is the fitted spectra. The double Gaussian fitting reduces the \chisq~value to 0.87. Panel L shows the H $\alpha$ profile of the RBE (black solid line) overlaid on the quiet Sun reference profile (dashed line). This event has a clear RBE-like H $\alpha$ profile with strong absorption in the blue wing compared to the red wing. The best fit Si IV requires a second broadening component which is blue-shifted.

\subsection{An event showing broadening and line core absorption}

In addition to the broadening and line core absorption, Event 4 has strong absorption in the blue wing compared to the red wing. At the beginning of the event, there is a strong brightening in the SJI 1400 \AA\ channel that matches the orientation and shape of the event in H $\alpha$. The brightness of the feature reduces as time progresses but the shape and orientation of the bright feature continue to follow the H $\alpha$ RBE. Si~{\sc iv} spectral evolution of the event is presented in Figure \ref{fig_td_ev4}. The format of the figure is the same as in Figure \ref{fig_td_ev2}. It is clear from the figure, that the H $\alpha$ jet produces a spatio-temporal enhancement in intensity and line width of Si~{\sc iv}. Similar to the previous example, there is no evidence of blue shift associated with this event in the transition region. The location of the event appears to be mostly red-shifted. However, asymmetry in the blue wing, seen in the spectral profile of strongly broadened line could be an indication of a blue shift.

Event 4 shows line broadening in the H $\alpha$ spectral line. This line broadening is asymmetrical, with 81\% more absorption in the blue-wing compared to the red-wing. This broadening is seen in the Si~{\sc iv} spectral profiles. This indicates that such events are complex RBEs. However, they reach transition region temperatures and are responsible for some of the brightenings in the transition region. The best fit for Si~{\sc iv} implies a second broadened component, but no line-shifts.

\subsection{Events showing enhanced line core, with a strong blue-shift.}
\begin{figure*}
\centering
\includegraphics[width=0.98\textwidth]{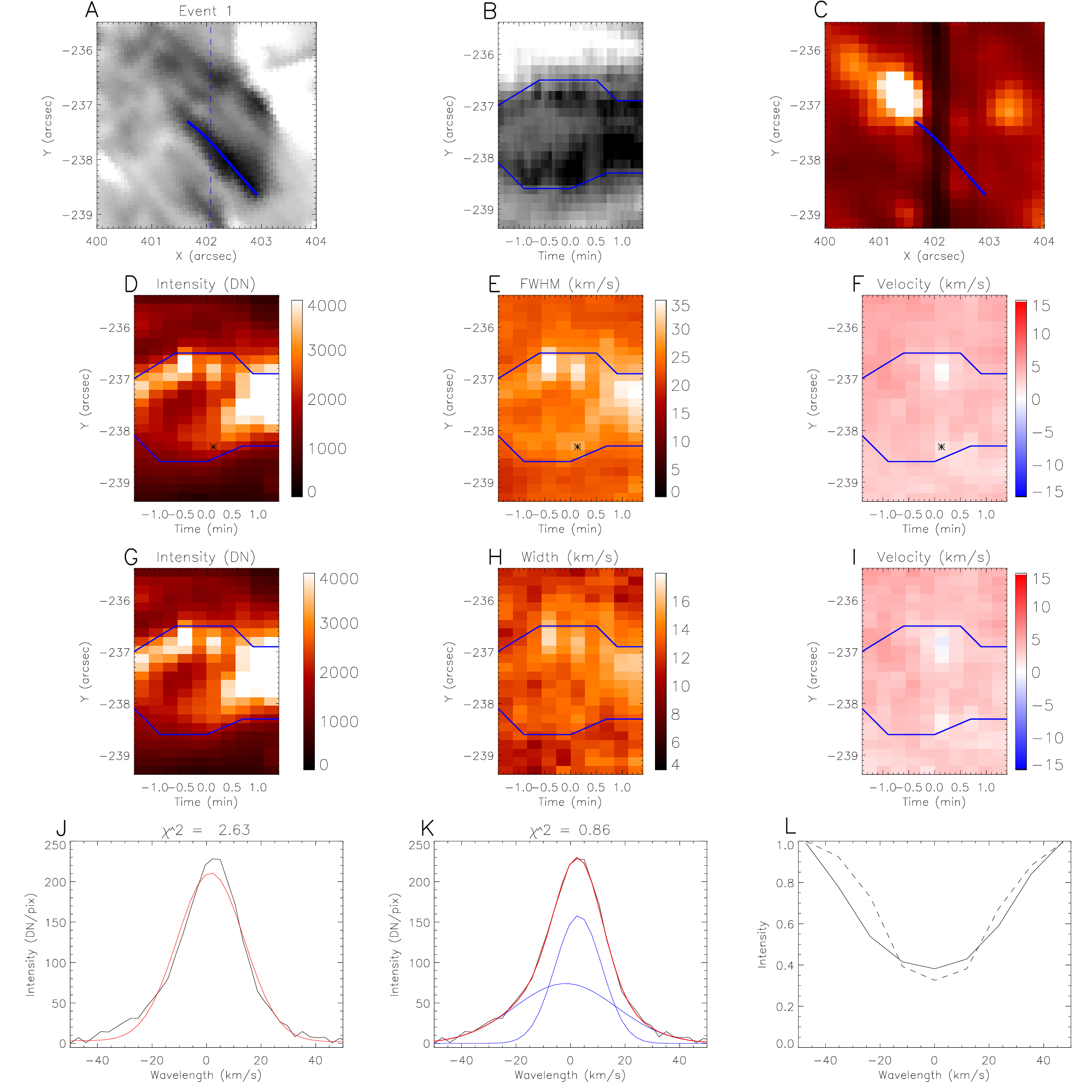}
\caption{Same as Figure \ref{fig_td_ev2} but for event 1. \label{fig_td_ev1}}
\end{figure*}
Events 1 and 3, are events that show broadening in the  H $\alpha$ line core, with a strong blue-shift. This line-core emission suggests that these events are yet another class of RBEs, as many RBEs show only a shift in the wings of the H $\alpha$ line and no intensity increase in the line-core. The Si~{\sc iv} 1394 \AA\ spectral analysis of event 1 is shown in Figure \ref{fig_td_ev1}. By comparing the time-distance plots of H $\alpha$ and Si~{\sc iv} 1394 \AA, it is clear that event 1 has a transition region counterpart that causes broadening and an intensity enhancement in the Si~{\sc iv} 1394 \AA\ line. The observed Si~{\sc iv} 1394 \AA\ profiles (panel J, K) of this event are asymmetrical with more emission in the blue wing compared to the red wing.

\subsection{Events showing broadening with stronger absorption in blue wing compared to red wing}

\begin{figure*}
\centering
\includegraphics[width=0.98\textwidth]{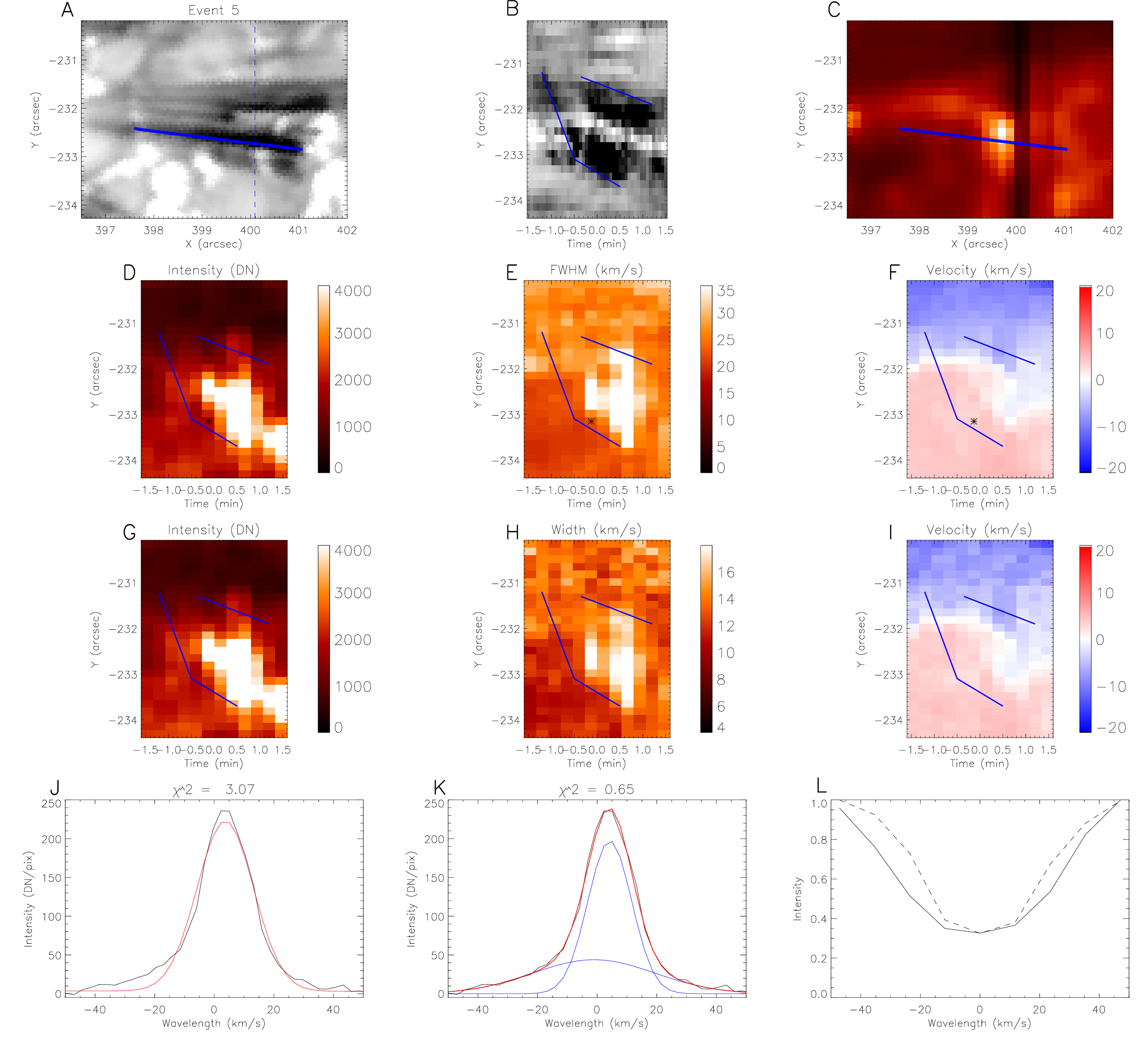}
\caption{Same as Figure \ref{fig_td_ev2} but for event 5. \label{fig_td_ev5}}
\end{figure*}

Event 5 in Figure \ref{fig_spectra} is an example of an RBE that shows broadening along with blue wing absorption. These events do not show any enhancement or reduction in core intensity compared to the reference quiet Sun profile. The time evolution of Si iv 1394 {\AA} spectral properties of event 5 is presented in Figure \ref{fig_td_ev5}. The time evolution plot in H $\alpha$ appears complex due its multi-threaded nature. However, there is clear signs of intensity enhancement (panel D, G) and broadening (panel E, H) in Si {\sc iv} 1394 \AA\ line at the location corresponding to the blue wing absorption in H $\alpha$ (black region in panel B). The Si {\sc iv} spectra for the bright region (pixel marked with x in panel D, E, F) is shown in panel J. The single Gaussian fit to the spectra yield a large value of ${\chi}^{2}$ due to the asymmetric broadening in the blue wing of Si {\sc iv }. A second Gaussian component was required to fit the spectra.

Our investigation of the transition region counterpart of H $\alpha$ reveals that there are different types of jet-like features in the chromosphere which can reach up to transition region temperatures, producing broadening and brightening in the Si~{\sc iv} 1394 \AA\ resonance line. Event 2 studied here is a clear example of an RBE producing a signature in the transition region. Even though events 1 and 4 cannot be considered as classical RBEs, they still produce brightenings and spectral line broadening in transition region lines. Moreover, there is evidence that most of these events show asymmetry in the Si~{\sc iv} 1394 \AA\ spectral line. This asymmetry in the Si~{\sc iv} profiles of both RBEs and RREs has been previously reported in \cite{Luc_2015_iris}. 

\begin{figure}
\centering
\includegraphics[width=0.48\textwidth]{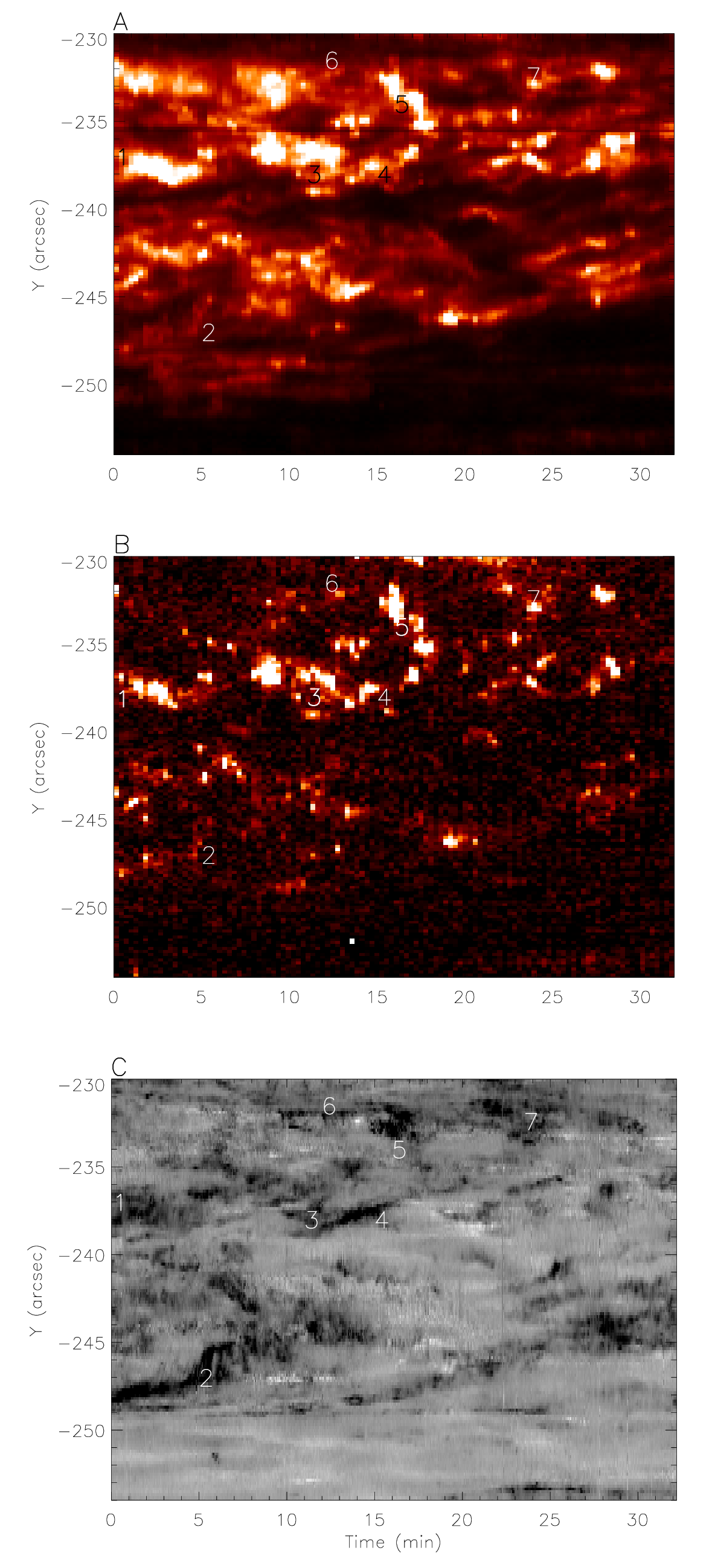}
\caption{Panel (a): Si {\sc iv} 1394 {\AA} intensity raster. Panel (b): Si {\sc iv} intensity raster at 1393.587 {\AA} ($\sim$ --36 km s$^{-1}$).
Panel (b): Constructed raster of H $\alpha$ blue wing at --35 km s$^{-1}$. The overlaid event numbers show the location of 7 RBEs that have a co-evolving Si~{\sc iv} spectral signature. Time 0 corresponds to 10-Jun-2014 07:50:53.320 UT.} 
\label{fig_all}
\end{figure}

In Figure \ref{fig_all}, we present the IRIS Si {\sc iv} 1394 {\AA} intensity rasters and its equivalent H $\alpha$ --35 km s$^{-1}$ blue wing observation. The time axis and the spatial length axis corresponds 
to the spatio-temporal window used for the RBE detection. Panel A shows the Si {\sc iv} 1394 {\AA} integrated intensity raster while the panel B displays the Si {\sc iv} intensity raster at 1393.587 {\AA}. The constructed  H $\alpha$ raster in panel C shows what IRIS would see if the sit-and-stare mode observation was done on the SST map. The overlaid numbers on the H $\alpha$ image marks the locations of the RBEs with co-evolving Si~{\sc iv} signature. The reduction in intensity of the H $\alpha$ --35 km s$^{-1}$ blue wing comes from features that show broadening or blue-shift of the H $\alpha$ spectral line. Therefore, in the observed region, reduction in the H $\alpha$ --35 km s$^{-1}$ blue wing intensity can be attributed to the spicule. There is an excellent match between the locations of the spicule and brightness observed in the integrated Si~{\sc iv} 1394 \AA\ intensity and in Si {\sc iv} blue wing image. This implies there is an association between spicules and transition region bright points. The plethora of chromospheric jets that reach transition region temperature may indicate that some of the transition region dynamics may be governed by the chromosphere below.

\section{Conclusions and discussions} \label{dis}

By studying the spicules in IRIS passbands, \cite{Pereira_2014} found that the spicules continue to evolve in IRIS passbands after they disappear from the cooler chromospheric Ca {\sc ii} H passbands. They interpret this result as a heating signature associated with spicules at least to transition region temperatures and conclude that the Ca {\sc ii} H Type II spicules are the initial stages of hotter events. Follow up work by \cite{skogsrud_2015} found visible signature of spicules in Ca {\sc ii} H, Mg {\sc ii} 2976\AA, Si {\sc iv} 1400 \AA, C {\sc ii} 1330 \AA, and He {\sc ii} 304 \AA\ filtergrams when observed off-limb. They were able to show that Ca {\sc ii} spicules evolve similar to the Mg {\sc ii} spicules as expected from the formation temperature of the lines. In their studies, the time evolution of spicules in Si {\sc iv} followed 2 distinct behaviour. In the first case, more than 50\% of the detected spicules showed an enhancement in Si {\sc iv} emission after the Ca {\sc ii} component fades. In the second case, Si {\sc iv} spicules appeared brighter at the top of the spicule which could indicate that the spicules are heated at least to transition region temperatures. 

The observed properties in H $\alpha$ and Si {\sc iv} of event 4 matches very well with the description of dynamic fibrils studied in \cite{Skogsrud_2016}. Using H $\alpha$, Ca {\sc ii} 8542 and Si {\sc iv} observations, \cite{Skogsrud_2016} found a tight correspondence between active region grains and dynamic fibrils. Their study shows grain emission is strongest during the initial ascending stage and it gets fainter in the later phase. The event studied here also shows a similar trend in the time evolution plot (see Figure \ref{fig_td_ev4}). Moreover, they found statistically significant enhancements in the line width of the Si {\sc iv} 1394/1403 \AA\ doublet and interpreted them as a signature of magneto-acoustic waves from below. Furthermore, they also found Si {\sc iv} 1403 \AA\ blue shift at the initial stage of the brightening. In some instances, a broadened Si {\sc iv} 1394 \AA\ line profile clearly shows evidence of blue-ward asymmetry. 

Based on the OCCULT algorithm we detected 1936 RBEs in an active region plage,  see Figure \ref{fig_chart}. Out of these 1936 events, only 90 events with RBE characteristics were covered by the narrow IRIS slit. 

In addition, we required that events should have a lifetime of at least 2 IRIS slit cadence ($\sim$32 s). These additional conditions were satisfied by 35 events. Of these 35 events, 14 of them were close to a large-scale active region flow. The Si {\sc iv} spectra associated with this region were dominated by a strong blue-shift for the entire duration of the observations. Therefore, we discarded those events which were close to the flow. For counting purposes, events with multiple threads would counted as one event.  
This reduced the event count under the IRIS to 11. Out of these 9 RBEs, only 2 of them show a standard blue-shifted H $\alpha$ profile as in \cite{Langangen_2008, sekse_2012}. The remainder has significant broadening in the red wing.  As shown earlier in Figure \ref{fig_chart}, the left side of the flow-chart of the detection scheme shows the results based on the SJI 1400 {\AA} channel, while the right side of the chart summarises the current study.

The present work showed that by comparing the spectral properties of RBEs across different spectral lines, covering different regions in the solar atmosphere, there are a diverse population of chromospheric jets that can reach up to transition region temperatures and produce enhancement in the Si {\sc iv} transition region line. 

The RBE-like features (whether single or multi-threaded) can be sub-divided into several categories;  
events with pure blue shifted H $\alpha$ profile without any absorption in the red wing, broadened line profile with the absorption in the blue is stronger compared to the red wing, core absorption and another 2 show core emission. However, this work does not discuss events that suddenly disappear at a particular wavelength within the H $\alpha$ profile, then appearing at a different wavelength, e.g. \cite{Judge_2011, Shetye2016, Pereira_2016}. Furthermore, \cite{Yur2020} using high cadence data from the Goode Solar Telescope operating at the Big Bear Solar Observatory, found H $\alpha$ RBEs that show very rapid morphological changes on timescales of the order of 1 s.

 Based on a cross-correlation of the RBE-like features as seen in H $\alpha$, just $\sim$50\% of them show an enhancement in the intensity of the IRIS 1400 \AA\ filter. Looking at the RBE-like events which have a Si {\sc iv} 1394 \AA\ line profile, 78\% of them show a Si {\sc iv} line flux increase. 
 From these, almost all of them (85\%) show a broadened blue-shifted Si {\sc iv} component at some time during their evolution.
The time when we see enhanced emission or line width indicates that a second broadened component is required which may or may not be blue-shifted. 
\cite{Luc_2015_iris} also showed that the Si~{\sc iv} profile has in some instance a second slightly blue-shifted broadened component.

\cite{Kleint2022} showed that only 0.01\% of all IRIS Si~{\sc iv} spectra show burst-like features. Although some of the Si~{\sc iv} spectra associated with the H $\alpha$ RBE reported here do show a weakly blue-shifted broadened component, none of them show the extensive dynamic spectra reported in the above paper. This is in agreement with work by \cite{Huang2017} who also looked at the same dataset as used here, but instead looked for highly broadened events termed explosive events; none were at the locations of the RBEs studied here. The present work  
shows that multiple spectral and imaging signatures dominate the accuracy of the detection methods used, implying that to detect the exact characteristics of spicules across the solar atmosphere needs careful analysis. We show the properties of RBEs and their signature in the transition region using co-spatial and co-temporal observations from CRISP/SST and IRIS. By studying the Si~{\sc iv} 1394 \AA\ spectra as well as the time series of the SJI 1400 \AA\ pass-band images, we find a clear signature of RBEs in the transition region.

\section{Acknowledgments}
The authors are most grateful to the staff of the SST for their invaluable support with the observations. The Swedish 1-m Solar Telescope is operated on the island of La Palma by the Institute for Solar Physics of the Royal Swedish Academy of Sciences in the Spanish Observatorio del Roque de los Muchachos of the Instituto de Astrofísica de Canarias. We would also like to thank the IRIS personnel for their help in planning these joint observations. Armagh Observatory \& Planetarium and NVN studentship is core funded by the N. Ireland Executive through the Dept for Communities. The authors wish to acknowledge the DJEI/DES/SFI/HEA Irish Centre for High-End Computing (ICHEC) for the provision of computing facilities and support. We also like to thank STFC and the Solarnet project which is supported by the European Commission’s FP7 Capacities Programme under Grant Agreement number 312495 for T\&S. JGD would like to thank the Leverhulme Trust for a Emeritus Fellowship. JS is funded by NSF grant no: 1936336.

\appendix
\section{Computing spectral properties by fitting IRIS spectra with Gaussians}\label{app}

The double Gaussian fitting was performed using the iris\_auto\_fit routine, which uses the Levenberg Marquardt algorithm. For the double Gaussian fitting, the fitting function is a linear combination of two single Gaussians. We used the template function to fit double Gaussian as suggested in \url{https://pyoung.org/quick_guides/iris_auto_fit.html}. The initial value of the centroid is approximately the rest wavelength of Si {\sc iv} 1394 \AA\ for both the Gaussian component and it is allowed to vary  $\pm 0.5$\AA\ from the centroid. Therefore, the algorithm can fit asymmetric components between $\pm 107$ km s$^{-1}$. The allowed width limit for the fit is between 0.02 to 0.55\AA\,. See \url{https://pyoung.org/quick_guides/iris_auto_fit.html} for more details. The above method gives a broad secondary component with a very small velocity shift in most of the jet-like events studied in this work, e.g. Figures 6 and 7. However, the same fitting method can also result in two Gaussian profiles which are largely separated in velocity. An example of a such profile is shown in Figure A1. The blue curves are the two Gaussian components and the red curve is the fitted spectra.

\begin{figure}
\centering
\includegraphics[width=0.4\textwidth]{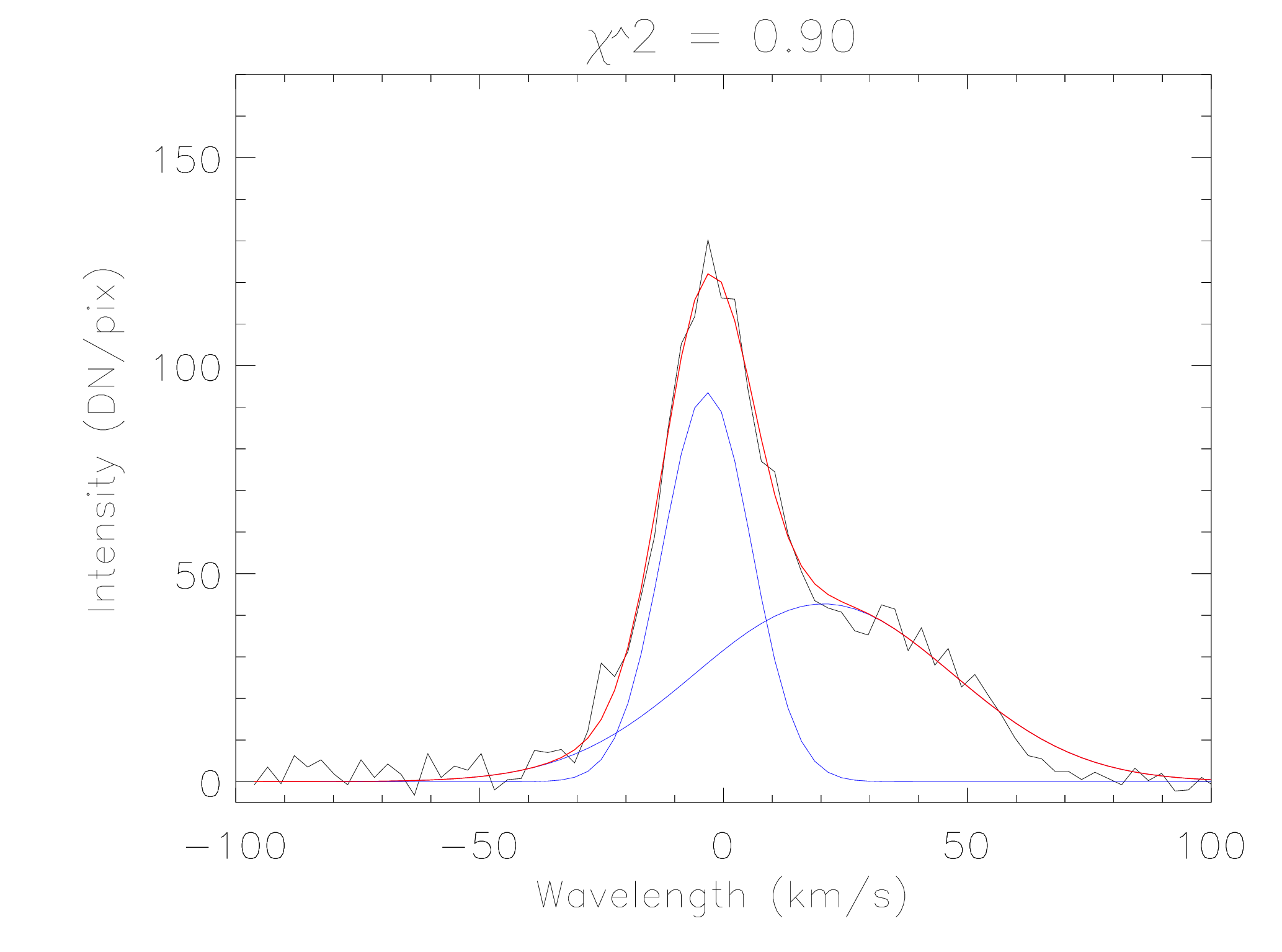}
\caption{An example of double Gaussian fit. \label{fig_app} }
\end{figure}

\section{Data Availability}
The reduced CRISP/SST data underlying this article will be shared on reasonable request to the corresponding author. All IRIS data is available via the IRIS webpage.
\bibliographystyle{mnras}
\bibliography{ref}

\end{document}